\documentclass{moriond}

\def\be{\begin{equation}}
\def\ee{\end{equation}}
\def\bea{\begin{eqnarray}}
\def\eea{\end{eqnarray}}

\newcommand{\Photo}{}
\newcommand{\nk}[1]{#1}
\usepackage{soul}

\begin{document}
\vspace*{4cm}
\title{THE LISA DATA CHALLENGES}

\author{Q. BAGHI$^1$, on behalf of the LDC Working Group}

\address{$^1$CEA Paris-Saclay University, Irfu/DPhP, Bat. 141, 91191 Gif sur Yvette Cedex, France}

\maketitle\abstracts{
The future space-based gravitational-wave detector LISA will deliver rich and information-dense data by listening to the milliHertz Universe. The measured time series will contain the imprint of tens of thousands of detectable Galactic binaries constantly emitting, tens of supermassive black hole merger events per year, tens of stellar-origin black holes, and possibly thousands of extreme mass-ratio inspirals. On top of that, we expect to detect the presence of stochastic gravitational wave backgrounds and bursts. Finding and characterizing many such sources is a vast and unsolved task. The LISA Data Challenges (LDCs) are an open and collaborative effort to tackle this exciting problem. A new simulated data set, nicknamed Sangria, has just been released with the purpose of tackling mild source confusion with idealized instrumental noise. This presentation will describe the LDC strategy, showcase the available datasets and analysis tools, and discuss future efforts to prepare LISA data analysis.
}

\section{Context}

The Laser Interferometer Space Antenna (LISA)~\cite{KarstenDanzmann2017} is scheduled for launch in 2034, and will explore the uncharted milliHertz frequency band of the gravitational-waves (GW) spectrum. This ESA-NASA mission of a new kind is planned to operate during 4 years, 
\nk{ with a possible extension}. A wealth of unprecedented science will flow from the analysis of its measurements, providing exceptional insights about the nature of gravity, the history of black holes, the evolution of compact stars in our Galaxy, and the very early history of the universe, to name a few. This science will only be enabled by a rigorous analysis of the data streams' content, which calls for novel processing techniques.

Unlike the monthly \nk{transient} events detected by the current terrestrial detectors LIGO-Virgo-KAGRA (LVK)~\cite{theligoscientificcollaboration2021gwtc3}, LISA will be sensitive to continuous sources mainly coming from the ultra-compact binaries (UCBs) in our Galaxy. These sources will remain in their inspiral phase during their observation, and constantly emit quasi-monochromatic waves. They are so numerous (several tens of millions~\cite{Lamberts2019}) that LISA can only  resolve a fraction of them, which still represents several tens of thousands of compact binary systems~\cite{Littenberg2020}. The farthest and less massive ones will create a confusion foreground which will stand above the instrumental noise between about 0.4 mHz and 4 mHz. Large sky transient surveys have already identified many UCB systems as LISA source candidates, called verification galactic binaries (VGBs)~\cite{Kupfer2018}. In addition, LISA will measure hundreds of transient GW signals. While the LVK network observes stellar-mass black hole binary mergers up to a few hundreds of solar masses, LISA will detect, for the first time, merging supermassive black holes (SMBH) with more than $10^5$ solar masses~\cite{Barausse2020}, likely testifying about the collision of distant galaxies. The signal-to-noise ratio (SNR) of these sources can reach a few thousands, one order of magnitude more than the loudest LVK detections. Yet, LISA might also detect stellar-mass black hole binaries in their early inspiral phase that LVK could detect later on, allowing for multiband detections. Beside\nk{s}, LISA should detect between 10 and 1000 sources of a very peculiar type, referred to as extreme-mass ratio inspirals (EMRIs)~\cite{Babak2017}. EMRIs are composed of a small \nk{compact object} orbiting a more massive black hole~\cite{Amaro_Seoane_2007}, with mass ratios larger than $10^4:1$. These systems will allow an in-depth understanding of general relativity and of the dynamics prevailing in dense star clusters. Beyond the listed GW emitters, there are more hypothetical and yet highly interesting sources LISA can reveal. Among them are stochastic gravitational wave backgrounds (SGWBs) coming from the amplification of the primordial quantum fluctuations in the early universe. Coming from well before the last scattering surface, the detection of a primordial SGWB would offer a unprecedented way to peer into the dark universe~\cite{Bartolo_2016}.

\section{Purpose of LISA data challenges}

The wealth of sources appearing in LISA measurements makes their extraction a unique challenge, which is a research subject of its own. The complexity of the data analysis comes from the high number, the long-lived nature, the diversity and the high mixing of the sources present in the observations. These features require special methodologies and strategies to perform the global fit, i.e., to detect and characterize all resolvable sources. Particularly, one needs to investigate what are the best data representations to use depending on the type of sources, but also how to combine different waveform models efficiently. An important aspect of this problem is parallelism, especially the conditions allowing one to simultaneously analyze separate segments of data (or separate domains of their representation) as well as different parts of the parameter space. To complexify further their processing, artefacts may corrupt some of the measurements. Besides the unavoidable presence of colored stochastic instrumental noise, spurious transients (or glitches), data gaps, spectral lines and non-stationarities will probably arise, which \nk{have to be taken} into account in the data modelling.

The purpose of the LISA Data Challenges (LDCs)\footnote{https://lisa-ldc.lal.in2p3.fr
} is to provide a framework to tackle various data analysis problems related to LISA, including the LISA global fit. 
LDCs are a recent resurrection of the historical mock data challenges (MLDCs)~\cite{Babak2010} which now fits into the structure of the LISA Consortium\footnote{https://www.elisascience.org/articles/lisa-consortium} as a science working group. The LDCs' goals are two-fold. First, they mean to provide a productive playground to stimulate the community's involvement in LISA data analysis research. Second, they serve as a support to validate LISA's science ground segment and assess the performance of prototype analysis pipelines.

\section{The LDC process}

\subsection{Measurement and pre-processing}

LISA forms a network of laser interferometers onboard three satellites forming a \nk{2.5} million kilometer-wide constellation \nk{following a} heliocentric orbit. Each spacecraft hosts two optical movable assemblies (MOSAs), each one including 3 interferometers. All interferometers yield light phase measurements, which can also be expressed as frequency variations with respect to the lasers' central frequencies. We call this quantity fractional frequency deviations. 

When a GW travels through the constellation, it changes the laser light travel paths between the satellites in a characteristic way. Each arm senses the projected space-time deformation along its length, which is recorded by the interferometers. Their data are downsampled to 4 Hz and then transmitted to Earth. Within the LISA distributed data processing center (DDPC), they will then be pre-processed through the interferometer noise reduction pipeline (INREP) to perform linear combinations of delayed measurements tailored to cancel the dominant noises. This process includes time-delay interferometry~\cite{tinto_time-delay_2020}, a technique designed to mitigate the largest noise, arising from the stochastic fluctuations of the laser frequencies. Prior to INREP processing, laser noise dominates the GW signal by 8 to 9 orders of magnitude. The main INREP data ouputs are called L1-level data and consist of the Michelson TDI variables $X, Y, Z$ which are the content of the LDC datasets. 

\begin{figure}
\centerline{\includegraphics[width=0.8\linewidth]{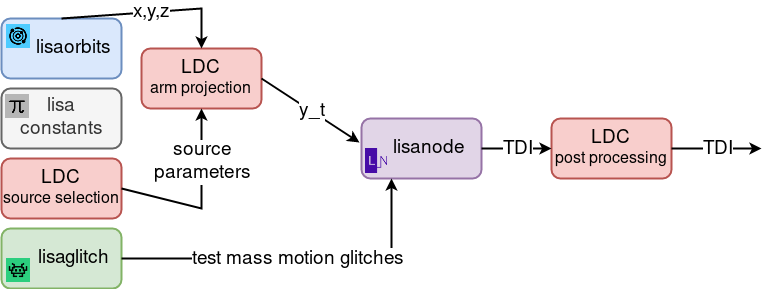}}
\caption[]{LDC data generation pipeline, including simulations tools like \texttt{LISA Orbits}~\cite{lisa_orbits_2022}, \texttt{LISA Glitch}~\cite{lisa_glitch_2022}, the end-to-end simulator \texttt{LISANode}~\cite{bayle2019} and the post-processing code \texttt{pyTDI}~\cite{pyTDI}.}
\label{fig:generation_pipeline}
\end{figure}

The LDC generation pipeline implements both the physics of the measurements and the TDI pre-processing up to simplifying hypotheses, as described in Fig.~\ref{fig:generation_pipeline}. 
This includes spacecraft orbital motion, projection of GWs onto the constellation arms, generation of noisy instrumental outputs, and realistic TDI processing. In addition, the pipeline can inject spurious transient phenomena called glitches which were observed in LISA Pathfinder~\cite{Baghi2022}.

\subsection{Publication of LDC simulated data sets}

The LDC designs challenges addressing specific or general issues, like assessing the ability to detect and characterize a particular GW source. Once a challenge is designed, the LDC then produces the associated dataset, i.e., simulated L1 data along with their metadata. Usually, the datasets come in two versions: a training version whose content is completely described by the metadata, including GW source parameters, and a blind version whose injected content remains undisclosed. After production, the LDC uploads the datasets on its website along with their documentation. The website is open to anyone willing to analyse the data, with a simple registration step to complete. 

The community then analyzes the published datasets with their own tools. Each challenge has a deadline associated to it, before which the participants can submit their results. 
Results may be in the form of parameter posterior samples or point estimates, and the LDC encourages labelling the parameters in the same way as the conventions adopted in the documentation. After submission, the LDC members compare the results qualitatively and quantitatively. If strong differences arises, they are reported to the participants who can investigate their analyses. \nk{After investigations are completed, the LDC group publishes the results of the challenge in a living review (see Section~\ref{sec:prospects}).}

\section{Ongoing challenges}

The new LISA Data Challenge working group has already released 3 main challenges, each one divided in sub-challenges. We list them below and describe their main characteristics. 

\subsection{\label{sec:radler}Challenge 1: Radler}
The LDC released the first challenge in 2019 under the code name \textit{Radler}, which refers to a beverage mixing beer and citrus juice, resulting in a low percent alcohol by volume that is supposed to convey the difficulty level of the data analysis. The \textit{Radler} challenge gathers 6 sub-challenges with only one GW source type per dataset, including a merging SMBH binary, an EMRI, ten VGBs, a simulated population of Galactic white dwarf binaries, a SGWB of primordial origin, and a few stellar-mass black hole binaries. This format allows participants to focus on source-specific data analysis methodologies. In addition, each challenge is fully specified, i.e., all injection parameters are available.

Many participants addressed the analysis of the SMBH (6 groups) and VGB (3 groups) sub-challenges and successfully retrieved the source parameters, usually using Bayesian stochastic search techniques like Markov chain Monte-Carlo (MCMC) algorithms~\cite{Katz2022,Cornish2020}. Four participants also tackled the SGWB sub-challenge, extracting the stochastic cosmological signal from the noise using template-based and template-agnostic techniques~\cite{Caprini_2019,Flauger_2021}. Two groups addressed the full Galaxy dataset analysis, with one complete inference using a trans-dimensional MCMC algorithm~\cite{Littenberg2020}. One group also successfully performed the parameter estimation of 22 stellar-mass black hole binaries~\cite{Buscicchio2021}, although the proper detection of this type of sources is not yet fully addressed. The EMRI sub-challenge \nk{was partially addressed although a careful examination of the submission remains to be done at this time.} \textit{Radler}'s official deadline has passed, but the LDC still accept new submissions.

\subsection{\label{sec:sangria}Challenge 2a: Sangria}
\begin{figure}[ht]
\centerline{\includegraphics[width=0.8\linewidth]{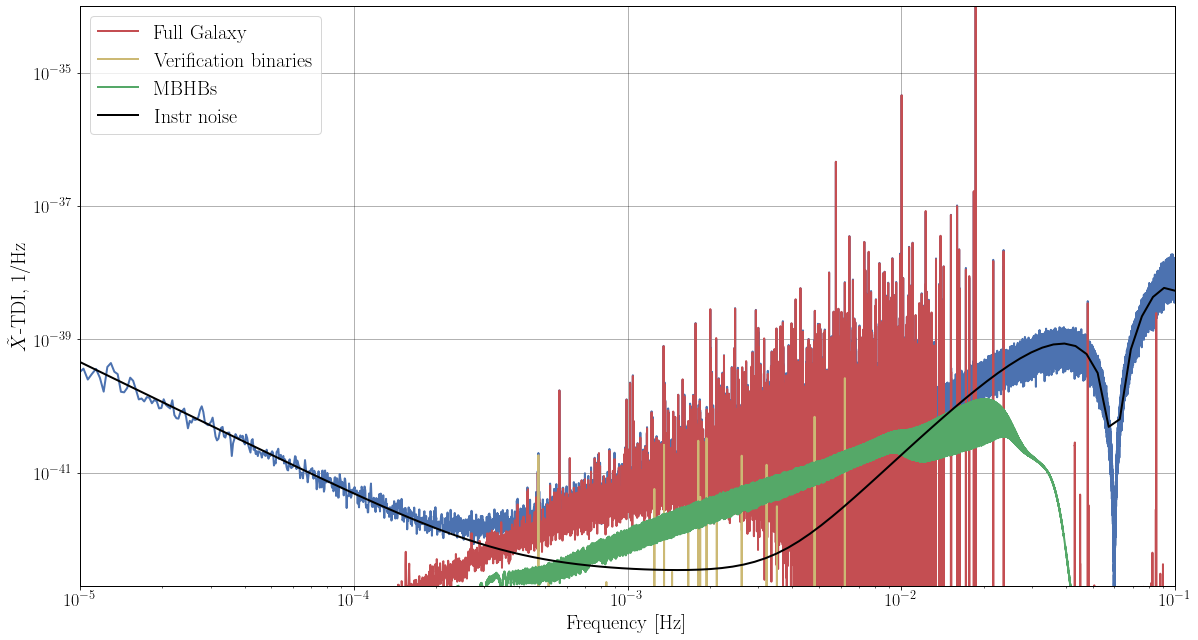}}
\caption[]{Fourier-transformed Michelson TDI channel X from the \textit{Sangria} simulated dataset, containing 30 millions of Galactic UCB sources (red) and 15 merging SMBH binaries (green). The simulation also features instrumental noise (black power spectral density).}
\label{fig:sangria}
\end{figure}

A new challenge was recently released with code name \textit{Sangria}, suggesting an increased difficulty with respect to the first challenge described in Sec.~\ref{sec:radler}. The goal is to foster the research on the global fit strategy and assess how massive black holes and Galactic binaries overlap. \textit{Sangria} contains two types of sources which are mixed together: a population of white dwarf binaries representing the Galaxy and 15 SMBH binary mergers (see Fig.~\ref{fig:sangria}). The simulation also includes instrumental noise. This challenge comes in two version: one with disclosed injected signal and noise parameters, the other completely blind. The \textit{Sangria} challenge is pending at the time of writing, with an official deadline set for October 1st, 2022.

\subsection{\label{sec:spritz}Challenge 2b: Spritz}
Another challenge called \textit{Spritz} was released shortly after \textit{Sangria}, to address another aspect of LISA data analysis, which is the presence of instrumental artefacts that we anticipate in the measurements. \textit{Spritz} includes three datasets, each one containing one GW source type hidden in data corrupted by instrumental glitches and data gaps. One dataset encloses a SMBH binary merger burried in instrumental noise affected by a translation of the loudest glitches observed in LISA Pathfinder. A similar dataset contains a SMBH binary merger with fainter and longer glitches. Finally, a third dataset features VGBs sources with several glitches drawn from a distribution representing LISA Pathfinder's population. All datasets are affected by data gaps whose pattern is defined by the current best estimate of the mission's measurement cycle. \textit{Spritz}'s deadline for submitting analysis results is the same as \textit{Sangria}'s (see Section \ref{sec:sangria}).

\section{\label{sec:prospects}Prospects}

The LDCs have set up a fruitful process and a framework to allow the community to work on key issues of LISA data analysis, with the objective of fostering new methodologies and advancing the research of science-extracting techniques for LISA. However, we are still a long way from being able to solve the global fit problem. The LDC working group members are currently writing the first issue of a living review picturing the current state of the art in LISA data analysis. New research outcomes enabled by the LDCs will then be appended along the years until the mission launches.

The goal we are setting for the current year is to complete the \textit{Sangria} challenge which mixes coalescing sources and continuous emissions from millions of compact binaries in the Milky Way, together with \nk{the \textit{Spritz} challenge which incorporates non-stationary instrumental perturbations}. This constitutes a major milestone that will be followed by other releases of more complex \textit{enchiladas}, blending more source types together. 
We will also pay a particular attention to expected sources that are known to be difficult to detect, such as stellar-mass black hole binaries and EMRIs. An upcoming challenge called \textit{Yorsh} will address the extraction of these specific signals. Regarding instrument simulation, future LDCs should feature more realistic noises including correlations and non-stationarities. 
Besides, we also plan to assess the flexibility and robustness of the developed methods, to allow for the discovery of new physics from LISA data.

\section*{Acknowledgments}

We acknowledge the support of the LISA Data Challenge working group from the LISA Consortium, especially Maude Lejeune who produced the plots and C\'{e}cile Cavet who maintains the LDC website.

\bibliography{references}

\end{document}